\documentclass
[notitlepage,12pt,groupedaddress,secnumarabic,prd,showpacs,showkeys,onecolumn,nofootinbib]{revtex4}%
\usepackage{amsfonts}
\usepackage{amsmath}
\usepackage{amssymb}
\usepackage{graphicx}
\usepackage{color}%
\begin{document}
\title{BPS equations and solutions for Maxwell-scalar theory}
\author{J.R. Morris}
\affiliation{Physics Department, Indiana University Northwest, 3400 Broadway, Gary, IN
46408, USA}
\email{jmorris@iun.edu}

\begin{abstract}
Energy minimizing BPS equations and solutions are obtained for a class of
models in Maxwell-scalar theory, where an abelian electric charge is immersed
in an effective dielectric of a real scalar field. The first order BPS
equations are developed using the straightforward \textquotedblleft on-shell
method\textquotedblright\ introduced by Atmaja and Ramadhan. Employment of an
auxiliary function of the scalar field allows a scalar potential that displays
a tachyonic instability. Consequently, a nontopological scalar soliton is
found to form around the charge. Examples and solutions are provided for (1) a
point charge or sphere in a flat Minkowski background, and (2) an
\textquotedblleft overcharged\textquotedblright\ compact object in a
Reissner-Nordstrom background. The solutions presented here for the former
(Minkowski) case recover those that have been previously obtained, while the
latter solutions are new BPS solutions.

\end{abstract}

\maketitle

\section{Introduction}

\ \ The possibility exists that elementary scalar fields may permeate the
universe. Such scalars could have many possible origins, such as those that
arise from scalar-tensor or tensor-multiscalar theories, extra dimensional
theories, or string theories. At any rate, it is important to investigate
possible interactions that scalar fields may have with known Standard Model
fields. These interactions may be of new, possibly noncanonical form. Cvetic
and Tseytlin \cite{Cvetic NPB94} have pointed out that a coupling of scalar
fields to the Maxwell field and kinetic terms of nonabelian gauge fields is a
generic feature of Kaluza-Klein and string theories. There, solutions have
been obtained for cases where only one scalar field contributes to the action,
as well as for cases where two scalar fields (dilaton and modulus) are
present. It is assumed that the flat spacetime solutions approximate the exact
solutions for the set of equations including gravity in regions where the
curvature is small. Other studies, such as those in \cite{BBFR 03}-\cite{Brito
EPJC14} have considered the effects of a dielectric medium, comprised of real
(or complex) \cite{Issifu AHEP21} scalar fields coupled to abelian and/or
nonabelian gauge fields. Such couplings permit new effects to emerge,
including those  that allow confining mechanisms.

\bigskip

\  Presently, attention is focused upon Maxwell-scalar theory, wherein a
neutral scalar field $\phi$ couples nonminimally to the Maxwell field
$F_{\mu\nu}$ via an \textquotedblleft effective permittivity\textquotedblright%
\ coupling function $\varepsilon(r,\phi)$, which here, is allowed to develop a
tachyonic instability that can result in a scalarization of an electric charge
immersed in this scalar field. These types of solitonic structures have been
recently investigated by Herdeiro, Oliveira, and Radu \cite{Herdeiro EPJC20}
and by Bazeia, Marques, and Menezes in \cite{Bazeia EPJC21} for the case of a
point charge immersed in a real scalar field, with the scalar-Maxwell
interaction described by the coupling $-\frac{1}{4}\varepsilon(\phi)F_{\mu\nu
}F^{\mu\nu}$. In those studies, energy minimizing first order Bogomolnyi
equations and solutions were obtained for a Minkowski background, describing
static, radially symmetric, charged solitonic objects. (See, e.g., Refs.
\cite{Bogol 76} and \cite{Prasad PRL75} for classical papers on the subject of
Bogomolnyi equations and \textquotedblleft BPS\textquotedblright\ solutions.)
Spherically symmetric static solutions for \textit{first order} (Bogomolnyi)
equations for the field $\phi(r)$, along with the electric field
$\mathbf{E}(r)$ were found by minimizing the system energy. The resulting
system describes a point charge surrounded by a scalar cloud. The presence of
the scalar cloud allows the description of a new type of soliton with electric
charge, and has an effect that modifies the electric field from that of
Maxwell theory without the nonminimal coupling.

\bigskip

\ \ The methods used in Refs.\cite{Herdeiro EPJC20} and \cite{Bazeia EPJC21}
work well for the case of spherical symmetry in a flat spacetime, where one
can perform first integrations to obtain Bogomolnyi equations. For example, in
Ref.\cite{Herdeiro EPJC20} a coordinate $x=1/r$ is defined, yielding a first
integral for the equation of motion for the scalar field $\phi$. One can then,
in principle, solve for $x(\phi)$, then invert to get $\phi(x)$ and transform
to get $\phi(r)$. With an assumed form for $\varepsilon(\phi)$, the functions
$\phi(r)$, $\varepsilon(r)$, and $\mathbf{E}(r)$ can be calculated. The
authors of Ref.\cite{Herdeiro EPJC20} then proceed by considering the
situation in a curved spacetime, and introduce a method of calculating the
functions $\phi(r)$, $\varepsilon(r)$, and $\mathbf{E}(r)$ with perturbation theory.

\bigskip

\ The method for obtaining Bogomolnyi equations used here appears to be more
direct, and follows that introduced by Atmaja and Ramadhan \cite{Atmaja
PRD14}, which in the present case, is quite simple to apply. Moreover, it can
be extended to accommodate the more general situation of a macroscopic compact
object in a Reissner-Nordstrom background, where gravitational effects are
taken into account. Also, \textit{exact closed-form} solutions can be
obtained, but at the cost of introducing a \textquotedblleft
noncanonical\textquotedblright\ coupling function $\varepsilon(r,\phi)$ (for
the case of a curved spacetime) which can carry an explicit dependence upon
the coordinate $r$. The function $\varepsilon(r,\phi)$ then depends on an
auxiliary function $X(\phi)$ (as in the Minkowski case), but also on the
spacetime metric $g_{\mu\nu}$. The idea is that the utility of a set of exact
solutions may illuminate the basic features expected in the Maxwell-scalar
theory with a coupling function of canonical form $\varepsilon(\phi)$. At any
rate, the previously obtained solutions for a Minkowski background can be
recovered in an economical way using the methods introduced here.

\bigskip

\ \ The idea of spontaneous scalarization of compact astrophysical objects,
such as neutron stars, was introduced by Damour and Esposito-Farese
\cite{Damour PRL93},\cite{Damour PRD96} within the context of scalar-tensor
theory, due to strong gravitational field effects associated with high
curvature. Much interest in the idea of scalarization has followed. (See, for
example, the review by Herdiero and Radu \cite{Herdeiro IJMPD15} and
references therein.) In particular, there has been interest in scalar fields
around charged compact objects. (Much literature exists on this topic, but see
Refs.\cite{Herdeiro EPJC20} and \cite{Bazeia EPJC21} and \cite{Herdeiro
PRL18}-\cite{Garcia 21} for example.)

\bigskip

\ As has been pointed out by Sotiriou \cite{Sotiriou CQG15}, there are several
motivations for being interested in scalar fields in curved spacetimes. These
include testing no hair theorems for black holes, constraining alternate
theories of gravity, and simply for the detection of scalar fields themselves,
for which black holes and possibly other compact objects may be a good
laboratory. Although compact objects may undergo a spontaneous scalarization
due to gravitational effects, the idea of scalarization can be extended to
situations where fields other than gravitation are involved.

\bigskip

\ \ Presently, Maxwell-scalar theory, as in \cite{Herdeiro EPJC20} and
\cite{Bazeia EPJC21}, is extended here to describe more general situations,
using the \textquotedblleft on-shell\ method\textquotedblright\ introduced by
Atmaja and Ramadhan \cite{Atmaja PRD14} to obtain Bogomolnyi equations and
solutions. (See also \cite{JM PRD21} and \cite{Mandal EPL21} regarding
solitons in curved spacetime.) With this method one introduces an auxiliary
function $X(\phi)$ that reduces the second order equation of motion for a
static, radially symmetric field $\phi(r)$, to a first order, energy
minimizing, Bogomolnyi equation, along with a constraint on the form of the
scalar potential $V(r,\phi)$, which depends upon $X(\phi)$ and the spacetime
metric. This method is applied to the case of (i) a point charge in Minkowski
spacetime, and (ii) a finite sized charged sphere, so that previously obtained
results found in \cite{Bazeia EPJC21} and for the toy model of \cite{Herdeiro
PRL18} and \cite{Herdeiro PRD21} are recovered. Next, the situation is
extended to describe a system comprised of a charged compact object within a
fixed external Reissner-Nordstrom (RN) background using the same forms of the
auxiliary functions, yielding new, \textit{exact analytical} solutions. For
these exact analytical solutions the stress-energy of the scalar field is
assumed to be negligible in comparison to that of the charged source and the
Maxwell field, so that any back reaction of the scalar field on the spacetime
geometry is ignored.

\bigskip

\ \ For microscopic systems with small values of mass and charge, the region
of space of interest is assumed to be at distances from a charge where the
metric is essentially flat, and the spacetime is well approximated by a four
dimensional Minkowski spacetime. However, on astrophysical scales, large
masses, and possibly large charges, may be encountered. It is speculated that
enormous charges $\sim10^{20}$ C might be supported by highly charged compact
stars, such as neutron stars \cite{Ray PRD03},\cite{Ray BJP04} or white dwarfs
\cite{Carvalho EPJC18}. (See also \cite{Pantop 21} and references therein
regarding highly charged compact objects.) For such cases, we assume a
Reissner-Nordstrom spacetime exterior to the compact object.

\bigskip

\ \ The Maxwell-scalar model is presented in Section 2, and in Section 3 a BPS
ansatz is given to obtain first order Bogomolnyi equations and exact analytic
solutions, based upon an assumed form of an \textquotedblleft effective
permittivity\textquotedblright\ function\ $\varepsilon(r,\phi)$ (which is
dictated by the choice of auxiliary function $X(\phi)$ and the spacetime
metric) that can display a tachyonic instability. The stress-energy and radial
stability of the BPS ansatz solutions are considered in Section 4. The BPS
ansatz solutions are determined to be radially stable. \ In Section 5,
scalarized charged objects are examined for (1) a point charge or charged
sphere in Minkowski spacetime, and (2) a charged compact object in a
Reissner-Nordstrom background. Exact solutions for the scalar and electric
fields are found for each case, with the former set of examples in flat
spacetime recovering those found in \cite{Bazeia EPJC21} and \cite{Herdeiro
PRL18}, with the corresponding cases for a Reissner-Nordstrom background
yielding \textit{new, exact, closed-form} nontopological solitonic BPS
solutions for the scalar field $\phi(r)$. A brief discussion forms Section 6.

\section{Maxwell-scalar model}

\ \ Consider the Maxwell-scalar (MS) model of a real scalar field $\phi$ that
is minimally coupled to gravity but nonminimally coupled to the Maxwell field
$F^{\mu\nu}$ via a coupling function $\varepsilon(r,\phi)$, where $r$ is the
radial coordinate for spherical symmetry. The geometry of the background
spacetime is described by a metric $g_{\mu\nu}$, and it is assumed that the
stress-energy of the scalar field $T_{\phi}^{\mu\nu}$ is negligible in
comparison to that of Maxwell fields and matter sources, so that scalar
backreactions on the spacetime can be ignored. We leave the exact nature of
the Maxwellian source unspecified for now, simply assuming it to be some
compact or point-like object with net (rationalized) electric charge
\cite{Note} $Q$.

\bigskip

\ The action to be considered for the region exterior to the source of charge
$Q$ in a fixed gravitational background is $S=\int d^{4}x\sqrt{g}%
\mathcal{L}(\phi,A_{\mu})$, with $\mathcal{L}$ being comprised of scalar and
Maxwell contributions, $\mathcal{L}=\mathcal{L}_{\phi}+\mathcal{L}_{EM}$. In
particular, we focus on cases where the magnetic field vanishes,
$\mathbf{B}=0$, but the electric field $\mathbf{E}(r)$ is sourced by some
charge $Q$. The spacetime metric is given by%
\begin{equation}
ds^{2}=A(r)dt^{2}-B(r)dr^{2}-r^{2}d\Omega^{2} \label{1}%
\end{equation}

where $d\Omega^{2}=d\theta^{2}+\sin^{2}\theta d\varphi^{2}$. We denote
$g=|\det g_{\mu\nu}|=A(r)B(r)r^{4}\sin^{2}\theta$ with the radial portion of
$\sqrt{g}$ given by $f(r)\equiv\sqrt{A(r)B(r)}\ r^{2}$ so that $\sqrt
{g}=f(r)\sin\theta$. We consider time independent fields that are spherically
symmetric, being functions of only $r$. In particular, it is assumed that a
static electric field $\mathbf{E}(r)$ exists due to the source $Q$, and a
static scalar field $\phi(r)$ exists in a state of equilibrium. (We do not
consider the equilibration process, during which $\phi=\phi(r,t)$.)

\bigskip

\ \ The action for our model is $S=\int d^{4}x\sqrt{g}\mathcal{L}$ with a
Lagrangian given by%
\begin{equation}
\mathcal{L}=\frac{1}{2}\partial_{\mu}\phi\partial^{\mu}\phi-\frac{1}%
{4}\varepsilon(r,\phi)F_{\mu\nu}F^{\mu\nu}-J^{\nu}A_{\nu},\label{2}%
\end{equation}

where $\varepsilon(r,\phi)$ is a nonminimal coupling function which may
display a tachyonic instability that can depend upon the radial distance $r$
from the coordinate origin. We want to consider the spatial region exterior to
the source of charge $Q$, i.e., the region of space where $J^{\nu}%
\rightarrow0$.

\bigskip

\ \ The equations of motion following from $\mathcal{L}$ are given by%
\begin{equation}%
\begin{array}
[c]{cc}%
\nabla_{\mu}\nabla^{\mu}\phi+\frac{1}{4}(\partial_{\phi}\varepsilon)F_{\mu\nu
}F^{\mu\nu}+\partial_{\phi}J^{\nu}A_{\nu}=0\smallskip, & \\
\nabla_{\mu}(\varepsilon F^{\mu\nu})=J^{\nu}, &
\end{array}
\label{3}%
\end{equation}

where $\partial_{\phi}=\partial/\partial\phi$. Using $F_{\mu\nu}F^{\mu\nu
}=-2(\mathbf{E}^{2}-\mathbf{B}^{2})$ and setting $J^{\nu}=0$ and
$\mathbf{B}=0$, these reduce to%
\begin{equation}%
\begin{array}
[c]{cc}%
\nabla_{\mu}\nabla^{\mu}\phi-\frac{1}{2}(\partial_{\phi}\varepsilon
)\mathbf{E}^{2}=0,\smallskip & \\
\nabla_{\mu}(\varepsilon F^{\mu\nu})=0. &
\end{array}
\label{4}%
\end{equation}

Upon implementing the assumption of spherical symmetry, the Maxwell equation
reduces to $\nabla_{r}(\varepsilon F^{r0})=\frac{1}{\sqrt{g}}\partial
_{r}(\sqrt{g}\varepsilon F^{r0})=\frac{1}{f}\partial_{r}(f\varepsilon
F^{r0})=0,$ which is solved by%
\begin{equation}
F^{r0}(r,\phi)=\frac{Q}{f(r)\varepsilon(r,\phi)}\ \ \ \text{and\ \ }%
\ \ F_{r0}=g_{rr}g_{00}F^{r0}=-\frac{A(r)Q}{f(r)h(r)\varepsilon(r,\phi
)}\label{5}%
\end{equation}

where we define a \textquotedblleft rationalized charge\textquotedblright%
\ $Q=Q_{0}/4\pi$, with $Q_{0}$ representing the actual charge, and%
\begin{equation}
\mathbf{E}^{2}=-F_{r0}F^{r0}=\frac{g_{00}|g_{rr}|Q^{2}}{f^{2}\varepsilon^{2}%
}=\frac{AQ^{2}}{f^{2}h\varepsilon^{2}}\ ,\label{6}%
\end{equation}

where $A(r)=g_{00}$, $h(r)=|g^{rr}|=1/B(r)$, and $f(r)=\sqrt{A(r)B(r)}r^{2}$.
The equation of motion for $\phi(r)$ reduces to%
\begin{equation}
-\frac{1}{f}\partial_{r}[fh\partial_{r}\phi]+\frac{AQ^{2}}{2f^{2}h}%
(\partial_{\phi}\varepsilon^{-1})=0.\label{7}%
\end{equation}

Using (\ref{4}) and (\ref{7}) to rewrite the equation of motion for $\phi$ as
$\nabla_{\mu}\nabla^{\mu}\phi+\partial_{\phi}V_{E}(r,\phi)=0$, or
\begin{equation}
-\frac{1}{f}\partial_{r}[fh\partial_{r}\phi]+\partial_{\phi}V_{E}%
(r,\phi)=0,\label{8}%
\end{equation}

allows the definition of an effective potential%
\begin{equation}
V_{E}(r,\phi)\equiv\frac{1}{2}\varepsilon\mathbf{E}^{2}=\ \ \frac{AQ^{2}%
}{2f^{2}h}\varepsilon^{-1}\ (r,\phi) \label{9}%
\end{equation}

with $\varepsilon(r,\phi)$ to be specified. We have $\partial_{\phi}%
V_{E}(r,\phi)=\partial_{\phi}(\frac{1}{2}\varepsilon\mathbf{E}^{2}%
)=-\ \frac{AQ^{2}}{2f^{2}h}(\frac{1}{\varepsilon^{2}}\partial_{\phi
}\varepsilon)=-\frac{1}{2}(\partial_{\phi}\varepsilon)\mathbf{E}^{2}$, so that
we have an effective equation of motion for $\phi$ given by $\square
\phi+\partial_{\phi}V_{E}(r,\phi)=0$.

\section{BPS ansatz}

\ \ We focus upon the effective equation of motion (EoM) for the scalar field.
By (\ref{8}), we have%
\begin{equation}
\partial_{r}[f(r)h(r)\partial_{r}\phi]=f(r)\partial_{\phi}V_{E}(r,\phi)
\label{10}%
\end{equation}

with $V_{E}(r,\phi)$ given by (\ref{9}). The \textquotedblleft on
shell\textquotedblright\ method of Atmaja and Ramadhan \cite{Atmaja PRD14} can
be used to generate a first order Bogomolnyi equation by subtracting a term
$\partial_{r}X(\phi)$ from both sides of (\ref{10}):%
\begin{equation}
\partial_{r}[f(r)h(r)\partial_{r}\phi-X(\phi)]=f(r)\partial_{\phi}V_{E}%
(r,\phi)-\partial_{r}X(\phi) \label{11}%
\end{equation}

Equation (\ref{11}) is then solved by solutions to the set of equations
\begin{subequations}
\label{12}%
\begin{align}
f(r)h(r)\partial_{r}\phi &  =X(\phi)\ ,\label{12a}\\
\ f(r)\partial_{\phi}V_{E}(r,\phi) &  =\partial_{r}X(\phi)\ .\label{12b}%
\end{align}

The first equation is the first order Bogomolnyi equation, whose solution also
solves the second order EoM (\ref{8}), while the second equation expresses the
effective potential $V_{E}(r,\phi)$ in terms of $r$ and the function $X(\phi
)$, since%
\end{subequations}
\begin{equation}
\partial_{\phi}V_{E}=\frac{1}{f}\partial_{r}X=\frac{1}{f}\partial_{\phi}%
X\cdot\partial_{r}\phi=\partial_{\phi}X\cdot\frac{1}{f^{2}h}X=\frac{1}%
{2f^{2}h}\partial_{\phi}X^{2}\ .\label{13}%
\end{equation}

Integrating gives a potential $V_{E}=(2f^{2}h)^{-1}(X^{2}+c)$. Upon setting
the constant $c=0$ and requiring that the function $X(\phi)$ be chosen so that
$V_{E}$ is everywhere finite (outside the source $Q$) for a finite energy
solution, we have from (\ref{9})
\begin{subequations}
\label{14}%
\begin{align}
V_{E}(r,\phi) &  =\frac{AQ^{2}}{2f^{2}h}\varepsilon^{-1}=\frac{1}{2f^{2}%
h}X^{2}\ ,\label{14a}\\
\varepsilon^{-1}(r,\phi) &  =\frac{1}{AQ^{2}}X^{2}(\phi)\ .\label{14b}%
\end{align}

with $h(r)=1/B(r)$. From (\ref{5}),%
\end{subequations}
\begin{equation}
F^{r0}=\frac{Q}{f\varepsilon}=\frac{1}{AfQ}X^{2}\ ,\ \ \ F_{r0}=g_{rr}%
g_{00}F^{r0}=-\frac{1}{fhQ}X^{2}\label{15}%
\end{equation}

with the physical electric field%
\begin{equation}
E_{r}=\sqrt{g_{00}|g_{rr}|}F^{r0}=\frac{Q}{\varepsilon r^{2}}=\frac{1}%
{QAr^{2}}X^{2}\ ,\label{16}%
\end{equation}

where $f=\sqrt{AB}r^{2}$.

\bigskip

\ \ In summary, a solution $\phi(r)$ to the Bogomolnyi equation of
(\ref{12a}), and the corresponding effective potential $V_{E}(r,\phi)$ of
(\ref{12b}), are given by
\begin{subequations}
\label{17}%
\begin{align}
\int\frac{d\phi}{X(\phi)} &  =\int\frac{dr}{f(r)h(r)}=\int\frac{dr}{\sqrt
{A/B}r^{2}}\ \ ,\label{17a}\\
V_{E}(r,\phi) &  =\frac{1}{2f^{2}h}X^{2}=\frac{1}{2}\varepsilon\mathbf{E}%
^{2}=\frac{AQ^{2}}{2f^{2}h}\varepsilon^{-1}\ ,\label{17b}%
\end{align}

with an electric field (\ref{16}) and an \textquotedblleft effective
permittivity function\textquotedblright\ $\varepsilon(r,\phi)$ shown in
(\ref{14b}). The effective potential $V_{E}(r,\phi)$ has the form
$V_{E}(r,\phi)=F(r)P(\phi)$, where $F(r)=(f^{2}h)^{-1}$ and $P(\phi)=\frac
{1}{2}X^{2}(\phi)$. The function $X(\phi)$ determines the effective potential
$V_{E}(r,\phi)$, the coupling function $\varepsilon(r,\phi)$, and electric
field $\mathbf{E}(r,\phi)$. The BPS solution given by (\ref{17a}) determines
$\phi(r)$, which, when inserted into $X[\phi(r)]$ gives the functions
$\varepsilon$, $\mathbf{E}$, and $V_{E}$ as functions of $r$. Whether a
charged compact object will admit BPS scalarizing solutions or not depends
upon the form of $X(\phi)$ and possible tachyonic instabilities of the
effective potential $V_{E}(r,\phi)=\frac{1}{2}F(r)X^{2}(\phi)$. (Even in the
absence of BPS solutions to the first order Bogomolnyi equations, the full
second order equations of motion may exhibit non-BPS scalarizing solutions.)

\section{Stress-energy and stability}

\ \ The Lagrangian for systems under consideration, described by (\ref{2})
with $J^{\nu}=0$, can be written as $\mathcal{L}=\mathcal{L}_{1}(\partial
\phi)+\mathcal{L}_{2}(\phi,A^{\mu})$, where $\mathcal{L}_{1}=\frac{1}{2}%
g_{\mu\nu}\partial^{\mu}\phi\partial^{\nu}\phi$ and $\mathcal{L}_{2}=-\frac
{1}{4}\varepsilon(\phi)F_{\mu\nu}F^{\mu\nu}$\ \ The stress-energy for the
system is%
\end{subequations}
\begin{equation}%
\begin{array}
[c]{ll}%
T^{\alpha\beta} & =2\frac{\partial\mathcal{L}}{\partial g_{\alpha\beta}%
}-g^{\alpha\beta}\mathcal{L}=(2\frac{\partial\mathcal{L}_{1}}{\partial
g_{\alpha\beta}}-g^{\alpha\beta}\mathcal{L}_{1})+(2\frac{\partial
\mathcal{L}_{2}}{\partial g_{\alpha\beta}}-g^{\alpha\beta}\mathcal{L}_{2})\\
& =\partial^{\alpha}\phi\partial^{\beta}\phi-g^{\alpha\beta}(\frac{1}%
{2}\partial^{\mu}\phi\partial_{\mu}\phi)+\varepsilon\lbrack g^{\alpha\nu
}F_{\nu\gamma}F^{\gamma\beta}+\frac{1}{4}g^{\alpha\beta}F_{\gamma\nu}%
F^{\gamma\nu}]
\end{array}
\label{18}%
\end{equation}

which, for $\phi=\phi(r)$ and $\mathbf{B}=0$, yields energy density and stress
components%
\begin{equation}%
\begin{array}
[c]{ll}%
T^{00} & =-\frac{1}{2}g^{00}\partial^{r}\phi\partial_{r}\phi+\frac{1}{2}%
g^{00}\varepsilon\mathbf{E}^{2}\\
T^{rr} & =\frac{1}{2}(\partial^{r}\phi)^{2}+\frac{1}{2}g^{rr}\varepsilon
\mathbf{E}^{2}%
\end{array}
\label{19}%
\end{equation}

We define an energy density $\mathcal{H}\equiv T_{0}^{0}=\frac{1}{2}%
h(\partial_{r}\phi)^{2}+\frac{1}{2}\varepsilon\mathbf{E}^{2}=\frac{1}%
{2}h(\partial_{r}\phi)^{2}+V_{E}(r,\phi)$ so that by (\ref{9})
\begin{equation}
\mathcal{H}=T_{0}^{0}=\frac{1}{2}h(\partial_{r}\phi)^{2}+V_{E}(r,\phi
)=\frac{1}{2}h(\partial_{r}\phi)^{2}+\ \frac{1}{2f^{2}h}X^{2}(\phi) \label{20}%
\end{equation}

with a corresponding configuration energy%
\begin{equation}
E=\int d^{3}x\sqrt{g}T_{0}^{0}=4\pi\int\mathcal{H\ }f(r)dr \label{21}%
\end{equation}

with%
\begin{equation}
\frac{E}{4\pi}=\int\frac{1}{2}fh(\partial_{r}\phi)^{2}dr+\int X^{2}(\phi
)\frac{1}{2fh}dr \label{22}%
\end{equation}

We can note that the auxiliary function $X(\phi)$ can be identified with a
$\phi$ derivative of a superpotential $W(\phi)$, i.e., $X(\phi)=\pm
\partial_{\phi}W(\phi)$. Additionally, the first order Bogomolnyi equation
\begin{equation}
\partial_{r}\phi=\frac{1}{fh}X(\phi)=\pm\frac{1}{fh}\partial_{\phi}W(\phi)
\label{23}%
\end{equation}

provides a lower bound to the energy. This is seen in the following way. The
energy density of a system with an effective potential $V_{E}=\frac{1}%
{2f^{2}h}X^{2}$ is%
\begin{equation}%
\begin{array}
[c]{ll}%
\mathcal{H} & =T_{0}^{0}=\frac{1}{2}h(\partial_{r}\phi)^{2}+\frac{1}{2f^{2}%
h}X^{2}\\
& =\frac{1}{2}h\left[  (\partial_{r}\phi-\frac{1}{fh}X)^{2}+\frac{2}%
{fh}X\partial_{r}\phi-\frac{1}{f^{2}h^{2}}X^{2}\right]  +\frac{1}{2f^{2}%
h}X^{2}%
\end{array}
\label{24}%
\end{equation}

For ansatz solutions satisfying (\ref{23}), denoted by $\phi_{A}$, the
gradient and potential parts of $\mathcal{H}$ are connected by $\partial
_{r}\phi_{A}=\frac{1}{fh}X$, and for these ansatz solutions $\mathcal{H}%
(\phi)\rightarrow\mathcal{H}_{A}(\phi_{A})$, with $\mathcal{H}\geq
\mathcal{H}_{A}$, where $\mathcal{H}$ is the energy density of \textit{any}
solution to the second order equation of motion for $\phi$, and $\mathcal{H}%
_{A}$ is the energy density of an \textit{ansatz} solution $\phi_{A}$. That
is, $\phi_{A}$ provides a nontrivial solution with the lower bound
$\mathcal{H}_{A}$.
\begin{equation}
\mathcal{H}_{A}=\frac{1}{f}X\partial_{r}\phi_{A}-\frac{1}{2f^{2}h}X^{2}%
+\frac{1}{2f^{2}h}X^{2}=\frac{1}{f^{2}h}X^{2}=2V_{E} \label{25}%
\end{equation}

Identifying $X(\phi)=\pm\partial_{\phi}W(\phi)$,%
\begin{equation}%
\begin{array}
[c]{ll}%
\dfrac{E_{A}}{4\pi} & =%
{\displaystyle\int_{r_{0}}^{\infty}}
\dfrac{1}{f^{2}h}X^{2}fdr=\Big|%
{\displaystyle\int_{r_{0}}^{\infty}}
\dfrac{1}{fh}X\cdot\partial_{\phi}Wdr\Big|=\Big|%
{\displaystyle\int_{r_{0}}^{\infty}}
\partial_{\phi}W\cdot\partial_{r}\phi dr\Big|\\
& =\Big|%
{\displaystyle\int_{\phi(r_{0})}^{\phi(\infty)}}
\partial_{\phi}Wd\phi\Big|=\Big|W(\phi(\infty))-W(\phi(r_{0}%
))\Big|=\Big|\Delta W\Big|
\end{array}
\label{26}%
\end{equation}

\bigskip

\ \ By (\ref{17b}), $\frac{1}{2}\varepsilon\mathbf{E}^{2}=V_{E}=\frac
{1}{2f^{2}h}X^{2}$, so that with (\ref{19}), the radial stress for an ansatz
solution $\partial_{r}\phi=\frac{1}{fh}X$ is given by%
\begin{equation}
T_{r}^{r}=\frac{1}{2}g^{rr}(\partial_{r}\phi)^{2}+\frac{1}{2f^{2}h}%
X^{2}=0\ .\label{27}%
\end{equation}

Additionally, for BPS ansatz solutions, we have%
\begin{equation}
\partial_{r}\phi=\pm\sqrt{2|g_{rr}|V_{E}(r,\phi)}=\pm(fh)^{-1}X(\phi
)\ .\label{27a}%
\end{equation}

This is a radial generalization of the usual one-dimensional linear result
$\partial_{x}\phi=\pm\sqrt{2V(\phi)}=\pm\partial_{\phi}W(\phi)$, where
$W(\phi)$ is a superpotential, with $X$ playing the role of $\partial_{\phi}W$
(as well as a generalization of Eq.(3) of Ref.\cite{Bazeia PRL03}.) The radial
stress vanishes for an ansatz solution, indicating a radial stability against
spontaneous radial contraction or expansion. (Another argument for the radial
stability for this type of system is given in \cite{JM PRD21} using scaling
arguments along the lines of Derrick's theorem \cite{Derrick}.) This radially
stable scalar field configuration differs from that of an ordinary, unstable
scalar domain wall bubble with a canonical potential $V(\phi)$ in the absence
of any external electric field.

\section{Scalarized charged objects}

\ \ Consider now a charged source embedded within a scalar field, which acts
as an effective permittivity medium, with an \textquotedblleft effective
permittivity\textquotedblright\ $\varepsilon(r,\phi)$. This \textquotedblleft
permittivity\textquotedblright, or Maxwell-scalar \textit{coupling function},
may exhibit a tachyonic instability, allowing a solitonic scalar field to
develop in the space around the charge. We focus on a final state of
equilibrium where the scalar field $\phi(r,t)$ has settled down to a static,
radially symmetric configuration $\phi(r)$.

\bigskip

\ \ Two separate models for the form of the effective permittivity
$\varepsilon(r,\phi)$ are considered, each having been examined previously for
the special case of a Minkowski background. (See \cite{Bazeia EPJC21} and
\cite{Herdeiro PRL18}.) In this case, the coupling function $\varepsilon
(\phi)$ has no explicit dependence upon $r$. Those results for BPS solutions
$\phi(r)$ are recovered here, and then our method is extended for the case of
a Reissner-Nordstrom background. From (\ref{17b}) the scalar-Maxwell
interaction is associated with an effective potential $V_{E}(r,\phi)=\frac
{1}{2f^{2}h}X^{2}(\phi)=\frac{Q^{2}A}{2f^{2}h}\varepsilon^{-1}(r,\phi)$. The
form of $X(\phi)$, along with the background metric $g_{\mu\nu}$, then
determine the structure of the coupling function $\varepsilon(r,\phi)$ for the
admission of BPS solutions (see Eq. (\ref{14})). Our first model is based on
the function $X(\phi)=\lambda(\eta^{2}-\phi^{2})$ so that the potential
$V_{E}(r,\phi)\propto\lambda^{2}(\eta^{2}-\phi^{2})^{2}$ exhibits a $Z_{2}$
symmetry that can be spontaneously broken. The vacua are located by $\phi
=\pm\eta$ with a local maximum at $\phi=0$. The tachyonic instability in the
potential allows for a nontopological BPS scalar soliton to form, with the
same analytic structure as obtained in \cite{Bazeia EPJC21} \footnote{See Sec.
2.2.1 of \cite{Bazeia EPJC21}.}. For the second model we take $X(\phi
)=\lambda(\eta^{2}-\phi^{2})^{1/2}$, giving rise to $\varepsilon(\phi)$
similar in form to that used in the toy model presented in \cite{Herdeiro
PRL18}, with $V_{E}\propto\lambda^{2}(\eta^{2}-\phi^{2})$. In this case
$V_{E}$ is unbounded below as a function of $\phi$, with a maximum at $\phi
=0$. Nevertheless, a finite-valued BPS solution for $\phi(r)$ exists, provided
that there is a short distance cutoff (as is the case of a finite conducting
sphere as in \cite{Herdeiro PRL18} and \cite{Herdeiro PRD21}), reproducing the
previously obtained results.

\bigskip

\ \ First, we look at a charge $Q$ in a Minkowski spacetime for each model,
and secondly, we consider the situation for a charged compact object of charge
$Q$ in a Reissner-Nordstrom background, where the object's effects due to mass
and charge are not neglected. In particular, we will see that for a
Reissner-Nordstrom spacetime, where the scalarized source may be some
astrophysical object like a highly charged neutron star \cite{Ray
PRD03},\cite{Ray BJP04} or white dwarf \cite{Carvalho EPJC18}, we \ must focus
on the case of an \textit{overcharged} object to obtain a BPS solution. It is
seen that for the case of an undercharged object, or an extremal black hole,
our ansatz does not admit BPS solutions to first order equations, although
solitonic solutions to the full second order equations can exist. In each case
we assume a fixed spacetime background so that any backreaction of the scalar
field on the spacetime geometry can be ignored. In this sense, \textit{the
focus is upon the development of BPS equations and solutions for the scalar
fields and electric fields obtained,} rather than upon the structure of the
spacetime geometry due to backreactions. The resulting configurations describe
electrically charged nontopological solitonic objects of the type studied in
\cite{Bazeia EPJC21} and the toy model found in \cite{Herdeiro PRL18} and
\cite{Herdeiro PRD21}, which are then extended for the case of a curved
spacetime. We illustrate the emergence of \textit{exact, closed form} BPS
solutions for the two forms of the functions $X(\phi)$ and $\varepsilon
(r,\phi)$ considered here, leading to the recovery of previously obtained
solutions for a Minkowski background, but new solutions for a
Reissner-Nordstrom background.

\subsection{Charged body in Minkowski background}

\subsubsection{First model: $X(\phi)=\lambda(\eta^{2}-\phi^{2})$}

\ \ The case of a point charge was described in Ref.\cite{Bazeia EPJC21}, and
is briefly reproduced here to demonstrate the BPS ansatz presented
above\footnote{See Section 2.2.1 of Ref.\cite{Bazeia EPJC21}.}. In particular,
we consider an auxiliary function $X(\phi)$ leading to a permittivity function
$\varepsilon^{-1}(\phi)\propto\lambda^{2}(\eta^{2}-\phi^{2})^{2}$
corresponding to that used in Ref.\cite{Bazeia EPJC21}. For spherical
symmetry, the metric is $ds^{2}=dt^{2}-dr^{2}-r^{2}d\Omega^{2}$, so that for
this case%
\begin{equation}
A=B=h=1,\ \ \ \ f(r)=r^{2},\ \ \ \ \sqrt{g}=f(r)\sin\theta\ ,\label{28}%
\end{equation}

where $A=g_{00}$, $B=-g_{rr}$, and $h=|g^{rr}|=B^{-1}$.

\bigskip

\ \ A point charge $Q_{0}=4\pi Q$ is taken to be located at the coordinate
origin, and the equations of motion for the scalar and Maxwell fields are
given by (\ref{4}). The electric field components are given by (\ref{5}) and
(\ref{6}). Assuming spherical symmetry, the scalar field EoM is given by
(\ref{10}), where the effective potential is given by (\ref{9}),
$V_{E}=\ \ \frac{AQ^{2}}{2f^{2}h}\varepsilon^{-1}$. For the flat spacetime
$V_{E}\rightarrow Q^{2}/(2\varepsilon r^{4})$. For a suitably chosen auxiliary
function $X(\phi)$, the resulting Bogomolnyi solution for the scalar field
$\phi$ and the effective potential $V_{E}$ are given by (\ref{17}), which in
this case become%
\begin{equation}
\int\frac{d\phi}{X(\phi)}=\int\frac{dr}{r^{2}},\ \ \ \ V_{E}(r,\phi)=\frac
{1}{2r^{4}}X^{2}(\phi) \label{29}%
\end{equation}

By (\ref{16}) and (\ref{14b}), the electric field $E_{r}$ and permittivity
function $\varepsilon$ are given by%
\begin{equation}
E_{r}=\frac{Q}{\varepsilon r^{2}}=\frac{1}{Qr^{2}}X^{2},\ \ \ \ \varepsilon
^{-1}(\phi)=\frac{1}{Q^{2}}X^{2}(\phi) \label{30}%
\end{equation}

\ \ We proceed by choosing a simple auxiliary function which allows a
tachyonic instability for $\varepsilon$, specifically,%
\begin{equation}
X(\phi)=\lambda(\eta^{2}-\phi^{2})=K\eta(1-\psi^{2}),\ \ \ \ \ (K\equiv
\lambda\eta,\ \ \psi\equiv\phi/\eta) \label{31}%
\end{equation}

(Note: $K\eta$ and $\psi$ are dimensionless.) By (\ref{29}), (\ref{31}),
(\ref{30}), and (\ref{20}) we have a solution $\psi(r)$, auxiliary function
$X$, electric field $E_{r}(r)$, and energy density $\mathcal{H}=T_{0}%
^{0}=2V_{E}$, given as (Fig.1, Fig.2)

\bigskip%
\begin{subequations}
\label{32}%
\begin{align}
\psi(r) &  =\frac{\phi(r)}{\eta}=-\tanh\Big(\frac{K}{r}\Big),\label{32a}\\
X(r) &  =K\eta\ \text{sech}^{2}\Big(\frac{K}{r}\Big),\label{32b}\\
E_{r}(r) &  =\frac{1}{Qr^{2}}X^{2}=\frac{K^{2}\eta^{2}}{Qr^{2}}\ \text{sech}%
^{4}\Big(\frac{K}{r}\Big),\label{32c}\\
\mathcal{H}(r) &  =2V_{E}(r)=\frac{1}{r^{4}}X^{2}(r)=\frac{K^{2}\eta^{2}%
}{r^{4}}\ \text{sech}^{4}\Big(\frac{K}{r}\Big).\label{32d}%
\end{align}

\begin{figure}[tbh]
\centering
\includegraphics[width=8.0cm]{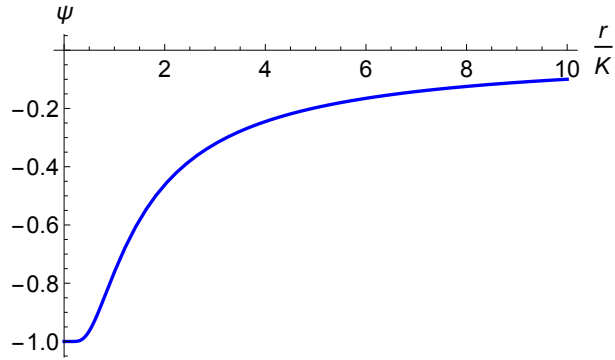}\caption{Sketch of the scalar field
solution of Eq.(\ref{32}), $\psi(r)=\phi(r)/\eta$ vs $r/K$, with $K\eta=1$. }%
\end{figure}

\begin{figure}[tbh]
\centering
\includegraphics[width=8.0cm]{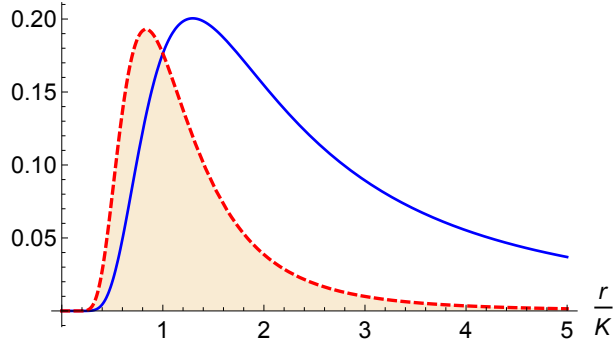}\caption{Sketches of the electric
field, $QE_{r}(r)$ (solid) and the field energy density, $\mathcal{H}(r)$
(dashed, shaded) \ of Eq.(\ref{32}) vs $r/K$. ($K\eta$ has been set to $1$.)}%
\end{figure}

\ \ For this model the BPS solution $\phi(r)$ is everywhere finite and
asymptotes to zero at $r=\infty$. The energy density $\mathcal{H}(r)$ is
localized, as typical for a solitonic configuration, and the electric field
$E_{r}(r)$ deviates substantially from a pure $1/r^{2}$ behavior, instead,
being pronounced with a maximum at a finite nonzero distance from the charge,
and approaching zero near the charge. The total energy (mass) of the scalar
field is%
\end{subequations}
\begin{equation}
\mathcal{M}=4\pi\int_{0}^{\infty}\mathcal{H}(r)r^{2}dr=4\pi K^{2}\eta^{2}%
\int_{0}^{\infty}\frac{1}{r^{2}}\text{sech}^{4}\Big(\frac{K}{r}\Big)dr=8\pi
K\eta^{2}\ .\label{mass1}%
\end{equation}

\ \ From (\ref{32a}) we have the asymptotic behavior $\phi(r)=-\eta
\tanh(K/r)\sim-K\eta/r$, so that we can identify a scalar charge $Q_{s}=\pm
K\eta$. (The $\pm$ sign is chosen since we could choose the $\pm$ sign for our
$X(\phi)$ to get the same $\mathbf{E}$ and $\mathcal{H}$.)

\subsubsection{Second model: $X(\phi)=\lambda(\eta^{2}-\phi^{2})^{1/2}$}

\ \ This type of system was briefly looked at as an example of scalarization
in the absence of gravity in Refs.\cite{Herdeiro PRL18} and \cite{Herdeiro
PRD21}. In this case, by (\ref{29}) and (\ref{30}), $\varepsilon^{-1}%
(\phi)\sim X^{2}(\phi)\sim\lambda^{2}(\eta^{2}-\phi^{2})$ and $V_{E}%
(r,\phi)=\frac{1}{2r^{4}}\lambda^{2}(\eta^{2}-\phi^{2})$. The effective
potential $V_{E}(r,\phi)\propto X^{2}(\phi)$ has no local minima in field
space, but the effective scalar field mass parameter $m_{\phi}^{2}$ is
negative, $\partial_{\phi}^{2}V_{E}(r,\phi)=-\lambda^{2}/r^{4}<0$, so that a
tachyonic instability presents itself.

\bigskip

\ \ Again, by (\ref{29}), (\ref{31}), (\ref{30}), and (\ref{20}) we have a
solution $\psi(r)=\phi(r)/\eta$, auxiliary function $X(r)$, electric field
$E_{r}(r)$, and energy density $\mathcal{H}(r)$ (Fig.3),
\begin{subequations}
\label{b1}%
\begin{align}
\psi(r) &  =\sin\Big(\frac{\lambda}{r}\Big),\label{b1a}\\
X(r) &  =\lambda\eta\cos\Big(\frac{\lambda}{r}\Big),\label{b1b}\\
E_{r}(r) &  =\frac{1}{Q^{2}r^{2}}X^{2}(r)=\frac{\lambda^{2}\eta^{2}}%
{Q^{2}r^{2}}\cos^{2}\Big(\frac{\lambda}{r}\Big),\label{b1c}\\
\mathcal{H}(r) &  =2V_{E}(r)=\frac{1}{r^{4}}X^{2}(r)=\frac{\lambda^{2}\eta
^{2}}{r^{4}}\cos^{2}\Big(\frac{\lambda}{r}\Big).\label{b1d}%
\end{align}

\begin{figure}[tbh]
\centering
\includegraphics[width=8.0cm]{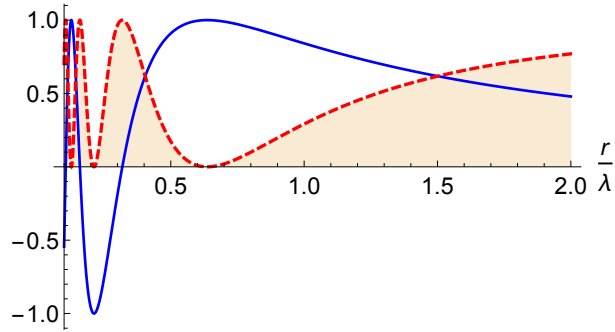}\caption{Sketch of the behavior of
the scalar field solution of Eq.(\ref{b1}), $\psi(r)=\phi(r)/\eta$ (solid) and
$X^{2}(r)$ (dashed, shaded) vs $r/\lambda$. \ The electric field $E_{r}(r)$
and energy density $\mathcal{H}(r)$ vary rapidly with large amplitudes for
small $r$.}%
\end{figure}

\ \ This BPS solution is also localized (provided that the radius $r_{0}$ of
the charge is nonzero), as are the electric field $E_{r}(r)$ and energy
density $\mathcal{H}(r)$. Again it can be noted that the electric field is
substantially different from a pure Coulombic case with a pure $1/r^{2}$
behavior. Instead, $E_{r}$ and $\mathcal{H}$ can oscillate at small distances
from the charge, approaching zero as $r\rightarrow\infty$. The oscillatory
behavior of $\mathcal{H}(r)$ can be associated with a sort of shell-like
structure near the charge. The total energy (mass) of the scalar field is%
\end{subequations}
\begin{equation}
\mathcal{M}(r_{0})=4\pi\int_{r_{0}}^{\infty}\mathcal{H}(r)r^{2}dr=4\pi
\lambda^{2}\eta^{2}\int_{r_{0}}^{\infty}\frac{1}{r^{2}}\cos^{2}\Big(\frac
{\lambda}{r}\Big)dr=\pi\lambda\eta^{2}\left[  \sin\Big(\frac{2\lambda}{r_{0}%
}\Big)+\frac{2\lambda}{r_{0}}\right]  \ ,\label{b2}%
\end{equation}

where $r_{0}$ is the radius of the charged object, serving as a short distance
cutoff. From (\ref{b1a}), asymptotically $\phi\sim\lambda\eta/r$, allowing an
identification of a scalar charge $Q_{s}=\pm\lambda\eta$.

\subsection{Charged body in Reissner-Nordstrom background}

\ \ The BPS ansatz for generating a first order Bogomolnyi equation for the
Maxwell-scalar system in a fixed spacetime background can be extended to the
case where an astrophysical object with mass $M$ and charge $Q_{0}=4\pi Q$ has
a non-negligible effect upon the spacetime geometry. Here, we consider the
case where the metric is the Reissner-Nordstrom metric:%
\begin{equation}
ds^{2}=\left(  1-\frac{r_{S}}{r}+\frac{r_{Q}^{2}}{r^{2}}\right)
dt^{2}-\left(  1-\frac{r_{S}}{r}+\frac{r_{Q}^{2}}{r^{2}}\right)  ^{-1}%
dr^{2}-r^{2}d\Omega^{2} \label{33}%
\end{equation}

with $r_{S}=2GM$ and $r_{Q}^{2}=GQ_{0}^{2}/4\pi=4\pi GQ^{2}$. This can be
rewritten in terms of geometrized mass and charge parameters $2\mu\equiv
r_{S}$ and $q\equiv r_{Q}$, with $\mu$ and $q$ both having units of length. We
then have%
\begin{equation}
ds^{2}=\left(  1-\frac{2\mu}{r}+\frac{q^{2}}{r^{2}}\right)  dt^{2}-\left(
1-\frac{2\mu}{r}+\frac{q^{2}}{r^{2}}\right)  ^{-1}dr^{2}-r^{2}d\Omega
^{2}\ .\label{34}%
\end{equation}

For this case we have%
\begin{equation}
A(r)=\left(  1-\frac{2\mu}{r}+\frac{q^{2}}{r^{2}}\right)  ,\ \ \ B(r)=A^{-1}%
(r),\ \ \ h(r)=A(r),\ \ \ f(r)=r^{2}.\label{35}%
\end{equation}

\ \ An important parameter to consider in this case is the ratio $\mu/q$ where
$q$ is the magnitude of the charge, although here, for definiteness, we
consider the charge to be positive. For $\mu>q$ there are two horizons located
by $r_{\pm}=\mu\pm\sqrt{\mu^{2}-q^{2}},$ and for the extremal state $\mu=q$
the two coincide. However, for $\mu<q$ there are no horizons. This is the case
that will be of interest here - the case $\mu/q<1$. The reason is that to find
a real valued first order Bogomolnyi solution using the BPS ansatz of
(\ref{17}), the integral on the right hand side of (\ref{17a}) must be real
valued for all $r>0$:%
\begin{equation}
\int\frac{dr}{f(r)h(r)}=\int\frac{dr}{\left(  1-\frac{2\mu}{r}+\frac{q^{2}%
}{r^{2}}\right)  r^{2}}=\frac{1}{\sqrt{q^{2}-\mu^{2}}}\tan^{-1}\left(
\frac{r/q-\mu/q}{\sqrt{1-\mu^{2}/q^{2}}}\right)  \ .\label{36}%
\end{equation}

For $\mu/q>1$ this becomes complex valued, and therefore no real valued
solution for all $r>\mu$ to the first order Bogomolnyi equation exists. We
therefore consider the case $q>\mu$. The possibility that this condition may
be physically realized for certain astrophysical objects\footnote{For example,
a highly charged neutron star with a mass on the order of $M\lesssim2$ solar
masses ($M\lesssim2M_{\odot}\lesssim2\times10^{57}$ GeV) and a net charge with
$Q_{0}\sim10^{20}$ C (i.e., $Q_{0}\sim10^{20}\times(2\times10^{18})\sim
2\times10^{38}$, natural units), one has $\mu/q=\sqrt{G}M/Q_{0}\lesssim1$.}
has been considered (see, for example,\cite{Ray PRD03},\cite{Ray
BJP04},\cite{Carvalho EPJC18}).

\subsubsection{First model: $X(\phi)=\lambda(\eta^{2}-\phi^{2})$}

\ \ Now we again take $X(\phi)=\lambda(\eta^{2}-\phi^{2})$ as given in
(\ref{31}), or $X=K\eta(1-\psi^{2})$, with $\psi=\phi/\eta$ and $K=\lambda
\eta$. Eqs. (\ref{17a}) and (\ref{36}) then yield the solution $\psi(r)$
(Fig.4),%
\begin{equation}
\psi(r)=\tanh\left\{  \frac{K/q}{\sqrt{1-\mu^{2}/q^{2}}}\Big[\tan
^{-1}\Big(\frac{r/q-\mu/q}{\sqrt{1-\mu^{2}/q^{2}}}\Big)\Big]\right\}
=\tanh\Theta(r),\label{37}%
\end{equation}

where%
\begin{equation}
\Theta(r)\equiv\frac{K/q}{\sqrt{1-\mu^{2}/q^{2}}}\Big[\tan^{-1}\Big(\frac
{r/q-\mu/q}{\sqrt{1-\mu^{2}/q^{2}}}\Big)\Big].\label{38}%
\end{equation}

With (\ref{37}), we obtain, with (\ref{14b}), (\ref{16}), and (\ref{25})
(Fig.4),
\begin{subequations}
\label{39}%
\begin{align}
X(r) &  =K\eta\left[  1-\tanh^{2}\Theta(r)\right]  =K\eta\ \text{sech}%
^{2}\Theta(r)\ ,\label{39a}\\
\varepsilon^{-1}(r,\phi) &  =\frac{1}{AQ^{2}}X^{2}(\phi)=\frac{K^{2}\eta
^{2}\ \text{sech}^{4}\Theta(r)}{Q^{2}(1-\frac{2\mu}{r}+\frac{q^{2}}{r^{2}}%
)}\ ,\label{39b}\\
E_{r} &  =\frac{Q}{\varepsilon r^{2}}=\frac{1}{QAr^{2}}X^{2}=\frac{K^{2}%
\eta^{2}\ \text{sech}^{4}\Theta(r)}{Qr^{2}(1-\frac{2\mu}{r}+\frac{q^{2}}%
{r^{2}})}\ ,\label{39c}\\
\mathcal{H} &  =T_{0}^{0}=2V_{E}=\frac{1}{f^{2}h}X^{2}(\phi)=\frac{K^{2}%
\eta^{2}\ \text{sech}^{4}\Theta(r)}{r^{4}(1-\frac{2\mu}{r}+\frac{q^{2}}{r^{2}%
})}\ .\label{39d}%
\end{align}

\begin{figure}[tbh]
\centering
\includegraphics[width=8.0cm]{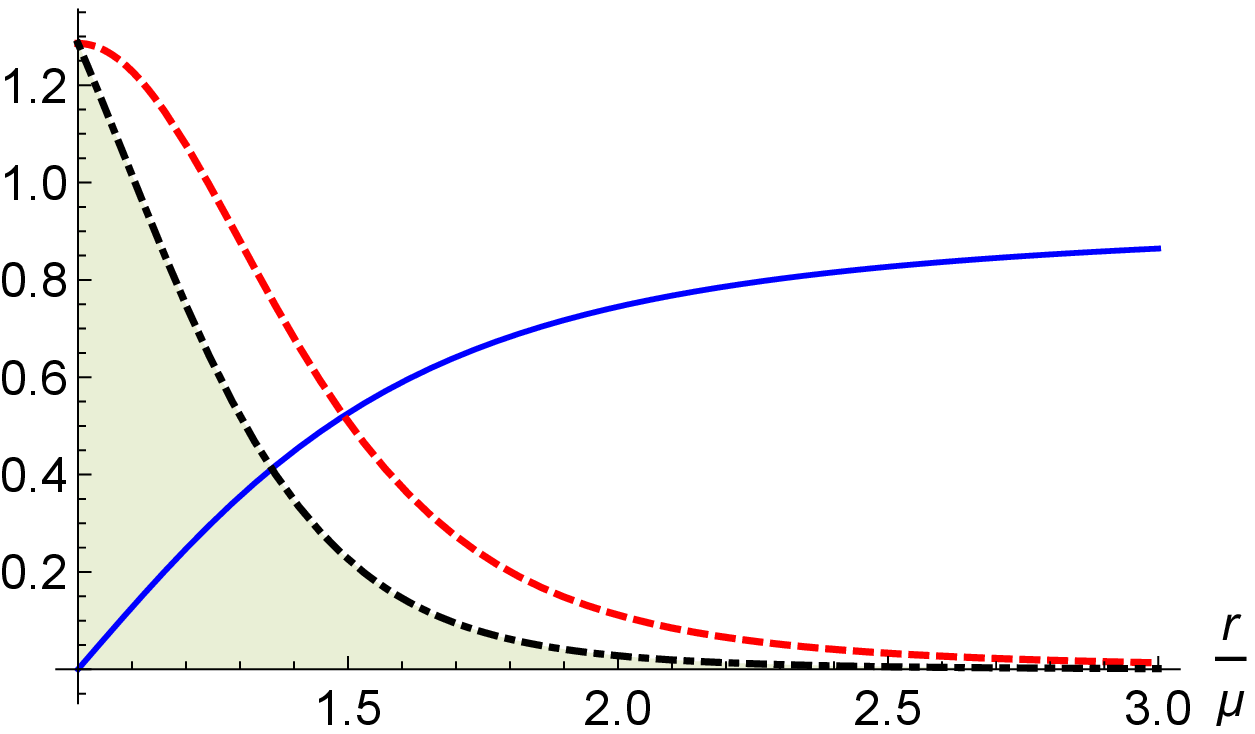}\caption{Sketches of the solution
$\psi(r)$ (solid) of Eq.(\ref{37}), electric field, $QE_{r}(r)$ (dashed) and
the field energy density, $\mathcal{H}(r)$ (dot-dashed, shaded) \ of
Eq.(\ref{39}) vs $r/\mu$. Here we use the settings $\mu/q=K/q=3/4$, and
$K\eta$ has been set to $1$.}%
\end{figure}

\ \ The solution for $\phi(r)$ is finite, and approaches a constant value of
$\phi(\infty)=\eta\tanh\Theta(\infty)=\eta\tanh[\frac{\pi K/q}{2\sqrt
{1-\mu^{2}/q^{2}}}]$. Again, the behaviors of $E_{r}(r)$ and $\mathcal{H}(r)$
differ from the Coulombic case, due to the factor of $A^{-1}(r)$%
sech$^{4}\Theta(r)$.

\bigskip

\ \ To get an expression for a scalar charge, we look at the asymptotic
behavior of (\ref{37}). First, we write (\ref{37}) as%
\end{subequations}
\begin{equation}
\psi(r)=\tanh[A\tan^{-1}(x)];\ \ \ A=\frac{K/q}{\sqrt{1-q^{2}/\mu^{2}}%
},\ \ \ x=\Big(\frac{r/q-\mu/q}{\sqrt{1-\mu^{2}/q^{2}}}\Big),\label{c1}%
\end{equation}

and
\begin{equation}
\tanh[A\tan^{-1}(x)]\sim\tanh[A(\pi/2-1/x)]\sim\tanh(\frac{\pi A}{2}%
)-\frac{A\text{sech}^{2}(\frac{\pi A}{2})}{x}\ .\label{c2}%
\end{equation}
As $r\rightarrow\infty$,
\begin{equation}
\frac{1}{x}=\sqrt{1-\mu^{2}/q^{2}}(\frac{r}{q}-\frac{\mu}{q})^{-1}\sim
\frac{q\sqrt{1-\mu^{2}/q^{2}}}{r}\ .\label{c3}%
\end{equation}

We then have $\phi(r)\sim\phi_{0}+Q_{s}/r$ as $r\rightarrow\infty$, where
$\phi_{0}=\phi(\infty)=\tanh(\frac{\pi A}{2})$. Then $\phi(r)\sim\phi_{0}%
-A$sech$^{2}(\frac{\pi A}{2})\cdot\frac{q\sqrt{1-\mu^{2}/q^{2}}}{r}$. A scalar
charge is then identified;%
\begin{equation}
Q_{s}=\pm\eta A\text{sech}^{2}(\frac{\pi A}{2})\cdot q\sqrt{1-\mu^{2}/q^{2}%
}=\pm K\eta\ \text{sech}^{2}(\frac{\pi A}{2})\ .\label{c4}%
\end{equation}

\subsubsection{Second model: $X(\phi)=\lambda(\eta^{2}-\phi^{2})^{1/2}$}

\ \ For this model we have the same integral given by (\ref{36}) and
(\ref{17a}) gives
\begin{equation}
\int\frac{d\phi}{X(\phi)}=\int\frac{d\phi}{\lambda(\eta^{2}-\phi^{2})^{1/2}%
}=\int\frac{d\psi}{\lambda(1-\psi^{2})^{1/2}}=\frac{1}{\lambda}\sin^{-1}%
\psi,\label{b3}%
\end{equation}
where $\psi=\phi/\eta$. Therefore, by (\ref{b3}) and (\ref{36}),%
\begin{equation}
\psi(r)=\sin\left\{  \frac{\lambda}{\sqrt{q^{2}-\mu^{2}}}\tan^{-1}\left(
\frac{r/q-\mu/q}{\sqrt{1-\mu^{2}/q^{2}}}\right)  \right\}  =\sin
\Phi(r),\label{b4}%
\end{equation}

where%
\begin{equation}
\Phi(r)\equiv\frac{\lambda}{\sqrt{q^{2}-\mu^{2}}}\tan^{-1}\left(
\frac{r/q-\mu/q}{\sqrt{1-\mu^{2}/q^{2}}}\right)  \ .\label{b5}%
\end{equation}

With (\ref{b4}) and with (\ref{14b}), (\ref{16}), and (\ref{25}) we have
(Fig.5)
\begin{subequations}
\label{b6}%
\begin{align}
X(r) &  =\lambda\eta(1-\psi^{2})^{1/2}=\lambda\eta\cos\Phi(r)\ ,\label{b6a}\\
\varepsilon^{-1}(r,\phi) &  =\frac{1}{AQ^{2}}X^{2}(\phi)=\frac{\lambda^{2}%
\eta^{2}\cos^{2}\Phi(r)}{Q^{2}(1-\frac{2\mu}{r}+\frac{q^{2}}{r^{2}}%
)}\ ,\label{b6b}\\
E_{r} &  =\frac{Q}{\varepsilon r^{2}}=\frac{1}{QAr^{2}}X^{2}=\frac{\lambda
^{2}\eta^{2}\cos^{2}\Phi(r)}{Qr^{2}(1-\frac{2\mu}{r}+\frac{q^{2}}{r^{2}}%
)}\ ,\label{b6c}\\
\mathcal{H} &  =T_{0}^{0}=2V_{E}=\frac{1}{f^{2}h}X^{2}(\phi)=\frac{\lambda
^{2}\eta^{2}\cos^{2}\Phi(r)}{r^{4}(1-\frac{2\mu}{r}+\frac{q^{2}}{r^{2}}%
)}\ .\label{b6d}%
\end{align}

\begin{figure}[tbh]
\centering
\includegraphics[width=8.0cm]{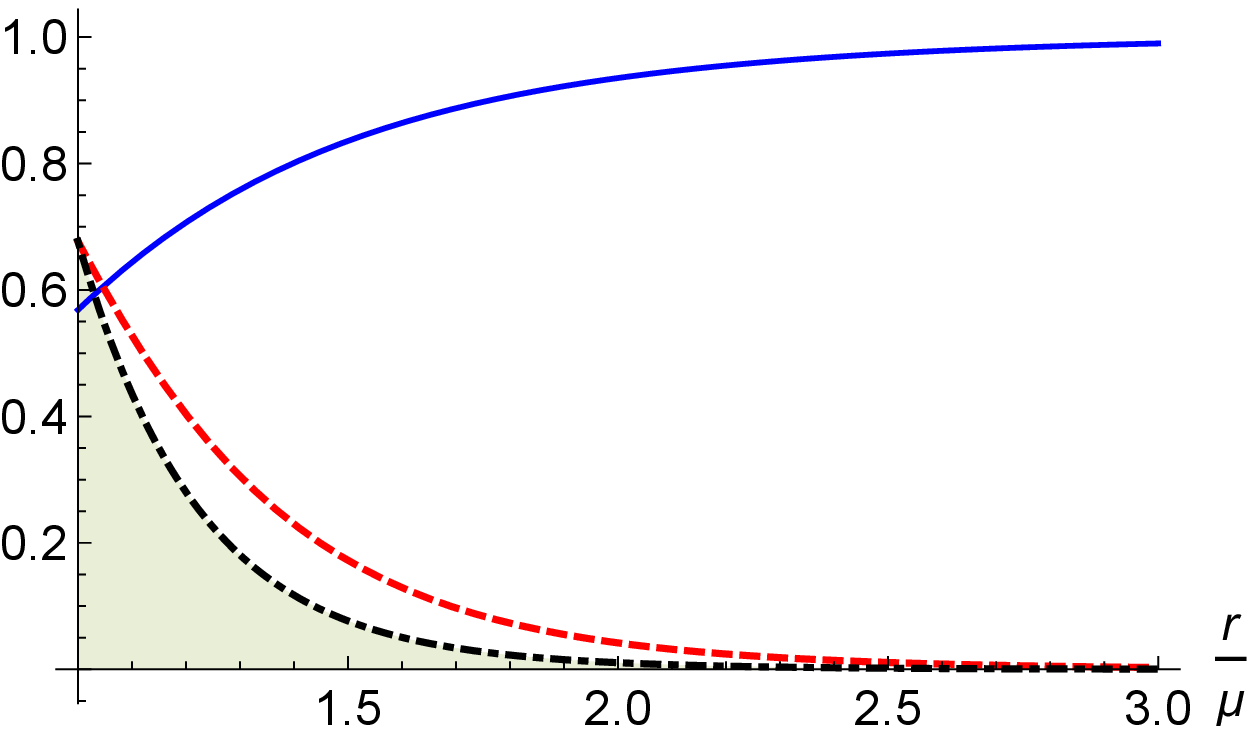}\caption{Sketches of the solution
$\psi(r)$ (solid) of Eq.(\ref{b4}), electric field, $QE_{r}(r)$ (dashed) and
the field energy density, $\mathcal{H}(r)$ (dot-dashed, shaded) \ of
Eq.(\ref{b6}) vs $r/\mu$. Here we use the settings $\mu/q=1/2$, and
$\lambda/q$ has been set to $1$.}%
\end{figure}

\ \ The solution for $\phi(r)$ is finite, and approaches a constant value of
$\phi(\infty)=\eta\sin[\frac{\pi\lambda}{2\sqrt{q^{2}-\mu^{2}}}]$. Again, the
behaviors of $E_{r}(r)$ and $\mathcal{H}(r)$ differ from the Coulombic case,
due to the factor of $A^{-1}(r)$cos$^{2}\Phi(r)$.

\bigskip

\ \ To find a scalar charge, write (\ref{b4}) as%
\end{subequations}
\begin{equation}
\psi(r)=\sin[C\tan^{-1}(x)];\ \ \ C=\frac{\lambda/q}{\sqrt{1-\mu^{2}/q^{2}}%
},\ \ \ x=\Big(\frac{r/q-\mu/q}{\sqrt{1-\mu^{2}/q^{2}}}\Big)\ .\label{c5}%
\end{equation}

As $r\rightarrow\infty$,
\begin{equation}
\psi(r)\sim\sin[C(\frac{\pi}{2}-\frac{1}{x})]\sim\sin(\frac{\pi C}{2}%
)-\frac{C\cos(\frac{\pi C}{2})}{x}\sim\sin(\frac{\pi C}{2})-\frac
{Cq\sqrt{1-q^{2}/\mu^{2}}}{r}\cos(\frac{\pi C}{2})\ .\label{c6}%
\end{equation}

Then $\phi(r)=\eta\psi(r)$ with $\phi\sim\phi_{0}-Q_{s}/r$, where $\phi
_{0}=\eta\sin(\frac{\pi C}{2})$ and
\begin{equation}
Q_{s}=\pm\eta Cq\sqrt{1-q^{2}/\mu^{2}}\cos(\frac{\pi C}{2})=\pm\lambda\eta
\cos(\frac{\pi C}{2})\ .\label{c7}%
\end{equation}

\bigskip

\ \ To briefly summarize, we note that real-valued \textit{first order
Bogomolnyi} solutions (for all $r>0$) \textit{are not} available for the
Reissner-Nordstrom metric here for geometrized parameters\ $q\leq\mu$ (in
which case there are horizons), which excludes such scalarized BPS ansatz
solutions for undercharged or extremal Reissner-Nordstrom black holes.
Solutions to the \textit{second order} equations of motion \textit{can} exist,
though. However, real-valued first order Bogomolnyi solutions \textit{are}
available for compact objects for which $q>\mu$ (in which case there are no
horizons). One possibility for such a compact object is an overcharged neutron
star or white dwarf \cite{Ray PRD03},\cite{Ray BJP04},\cite{Carvalho EPJC18},
or an object that has collapsed to a naked singularity before losing all of
its charge \cite{Ray PRD03}.

\section{Discussion}

\ \ The possibility of a scalar field being found concentrated around a
gravitating object, such as a black hole, has been investigated extensively
since the emergence of the idea in the context of scalar-tensor theory
\cite{Damour PRL93},\cite{Damour PRD96}. (See the review by Herdeiro and Radu
\cite{Herdeiro IJMPD15}, and references therein.) In addition, there have been
investigations into the possibility of scalarizing solutions around
electrically charged sources (see Refs.\cite{Cvetic NPB94},\cite{BBFR
03}-\cite{Issifu AHEP21},\cite{Herdeiro EPJC20},\cite{Bazeia EPJC21}, and
\cite{Herdeiro PRL18}-\cite{Garcia 21} for example.) By adopting a fixed
background metric, some of the complications due to gravitation can be evaded
while capturing essential qualitative features \cite{Bazeia EPJC21},
\cite{Herdeiro PRL18}, \cite{Herdeiro PRD21} of the scalar field.

\bigskip

\ \ The idea of scalarization within the context of Maxwell-scalar theory has
been considered here, where the \textquotedblleft on-shell
method\textquotedblright\ of Atmaja and Ramadhan \cite{Atmaja PRD14} has been
used to obtain exact, analytic closed-form solutions to first order Bogomolnyi
equations, along with a constraint on the form of the effective potential. The
first order Bogomolnyi equation can then be solved by quadrature. These
minimal energy BPS solutions automatically satisfy the second order
Euler-Lagrange equations of motion, and are radially stable, i.e., stable
against spontaneous radial expansion or collapse \cite{JM PRD21},\cite{Mandal
EPL21}. A nonlinear coupling function, or \textquotedblleft effective
permittivity\textquotedblright\ $\varepsilon(r,\phi)$, may allow a tachyonic
instability, which then gives rise to a scalar cloud, in the form of a
nontopological soliton, around an electrically charged source.

\bigskip

\ \ Here, two examples of BPS solutions for a real scalar field $\phi(r)$
responding to an electrically charged source have been presented with two
forms of an effective scalar-Maxwell coupling function $\varepsilon(r,\phi)$,
generated by the auxiliary functions $X(\phi)$. These two forms for
$\varepsilon$ have been looked at previously in the context of a flat
Minkowski background \cite{Bazeia EPJC21},\cite{Herdeiro PRL18},\cite{Herdeiro
PRD21}, where gravitation plays no part. These solutions for a Minkowski
background have been recovered here, and then extended to the case of a
Reissner-Nordstrom background, where both gravitation and electromagnetism are
present, and are not ignored. \textit{Exact, closed form}, \textit{analytical}
solutions have been found for each situation. These solutions provide
descriptions for the profiles of the scalar field $\phi(r)$, the electric
field $E_{r}(r)$, and the scalar field energy density $\mathcal{H}%
(r)=T_{0}^{0}(r)$.

\bigskip

\ \ The situations for a point charge or charged sphere in Minkowski
spacetime, were briefly reviewed, and the basic results found in
Refs.\cite{Bazeia EPJC21} and \cite{Herdeiro PRL18} were recovered. Next,
using similar coupling functions, a nonrotating charged compact object with an
exterior Reissner-Nordstrom background was considered. Specifically, an
overcharged object with a geometrized charge to mass ratio $q/\mu>1$ was seen
to provide an example of scalarization that could be relevant to highly
charged neutron stars \cite{Ray PRD03},\cite{Ray BJP04} or white dwarfs
\cite{Carvalho EPJC18}.

\bigskip

\ \ In these examples closed form analytical solutions have been obtained to
describe the scalar cloud and attendant electric field $\mathbf{E}(r)$. The
examples for a Minkowski background have recovered the solutions previously
presented in Refs.\cite{Bazeia EPJC21} and \cite{Herdeiro PRL18}, and the
solutions for the Reissner-Nordstrom background are, to our knowledge, new.
The effective permittivity $\varepsilon(r,\phi)$ gives rise to electric fields
$\mathbf{E}(r)$ which deviate substantially from the Coulombic ones that would
be expected from pure Maxwell theory with $\varepsilon=1$ and no scalar
coupling. If a gravitating source is electrically charged, and the electric
field is modified from its Coulombic form, then there may be observable
effects on a surrounding medium. For instance, radiation emitted by an
accelerated charged particle in a such an altered electric field would differ
from that of a charged particle in a Coulombic electric field of an
unscalarized charged source. Consequently, it may be possible to infer the
presence of scalar fields near charged, gravitating bodies. The features
exhibited by exact, closed form BPS solutions are expected to also be seen, at
least in a qualitative way, in non-BPS solutions where scalar back reactions
are accounted for, or for a case where a coupling function $\varepsilon(\phi)$
is not of the BPS form. In addition, it could be of interest to see what the
effects of a scalar mass or an additional scalar potential $V(\phi)$ may be.

\begin{center}

\end{center}


\begin{thebibliography}{99}
\bibitem {Note}For notational convenience (to avoid writing explicit factors
of $4\pi$), we use a rationalized charge $Q\equiv Q_{0}/4\pi$, where $Q_{0}$
is the total charge associated with an ordinary Gauss' law and electric field
in Minkowski space, $\nabla_{\mu}(\varepsilon F^{\mu0})=j^{0}\rightarrow$
$\oint\varepsilon\mathbf{E}\cdot d\mathbf{S}=Q_{0}=4\pi Q,$\ with$\ E_{r}%
=Q_{0}/(4\pi\varepsilon r^{2})=Q/(\varepsilon r^{2})$. The free space
permittivity is $\varepsilon_{0}=1$, and in natural units ($\hbar=c=1$),
$\alpha=e^{2}/4\pi$. We have $g_{00}=\Big(1-\dfrac{2GM}{r^{2}}+\dfrac
{GQ_{0}^{2}}{4\pi r^{2}}\Big)$ for the Reissner-Nordstrom metric.
\end{thebibliography}
\end{document}